\documentclass[pra,aps,twocolumn,showpacs]{revtex4}
\usepackage{epsfig}
\usepackage{graphicx}
\newcommand{\ket}[1]{|#1\rangle}
\newcommand{\bra}[1]{\langle #1|}

\begin{document}
\title{Noncyclic mixed state phase in SU(2) polarimetry}
\author{Peter Larsson and Erik Sj\"oqvist\footnote{Electronic 
address: eriks@kvac.uu.se}}
\affiliation{Department of Quantum Chemistry, 
Uppsala University, Box 518, S-751 20 Sweden}
\begin{abstract} 
We demonstrate that Pancharatnam's relative phase for 
mixed spin$-\frac{1}{2}$ states in noncyclic SU(2) 
evolution can be measured polarimetrically. 
\end{abstract}
\pacs{03.65.Vf; 03.75.Be}
\maketitle
Uhlmann's early work \cite{uhlmann86} on phases for mixed quantal
states was probably the first to address the issue of holonomy for
density operators. More recently, mixed state phases were reconsidered
by Sj\"{o}qvist {\it et al.} \cite{sjoqvist00}, primarily to provide
an operationally well defined concept of Pancharatnam's relative phase
\cite{pancharatnam56} as well as of geometric phase for such states in
unitary evolution. The two phases in \cite{uhlmann86} and 
\cite{sjoqvist00} have been shown 
\cite{slater01a,slater01b,ericsson02} to correspond to generically 
different holonomy effects and the restriction to nondegenerate
density operators for the geometric phase in \cite{sjoqvist00} has
been removed in \cite{singh03}. Furthermore, \cite{sjoqvist00} has met
extension to the off-diagonal case \cite{filipp03} as well as to
nonunitary evolution \cite{ericsson03}, and an interferometric 
experimental study, using nuclear magnetic resonance technique, 
has been carried out \cite{du03}.

In this Letter, we demonstrate that the noncyclic mixed state phase in
\cite{sjoqvist00} may be tested in the spin$-\frac{1}{2}$ case using
the polarimetric setup proposed by Wagh and Rakhecha in \cite{wagh95}
and implemented experimentally for neutrons in \cite{wagh00}. Such an
experiment is potentially advantageous as polarimetry could offer
better precision and robustness compared to interferometry
\cite{rakhecha01}. We thus believe that the present analysis is useful
when it comes to high-precision test of the predictions in
\cite{sjoqvist00}.

Let us first briefly describe the Wagh-Rakhecha polarimetric setup
sketched in Fig. 1.  A $\pi/2$ flip around the $y$ axis, say, is
applied to a single beam of spin$-\frac{1}{2}$ particles chosen to be
polarised along the $+z$ direction, creating a superposition
of two orthogonal states. The components of the superposition acquire
opposite Pancharatnam phases under the influence of the unitarity
$U$. Another $-\pi/2$ rotation around the $y$ axis is followed by a
measurement of the output intensity that results from a projection
onto the $+z$ axis.

\begin{figure}[htbp]
\centering
\includegraphics[width=8 cm]{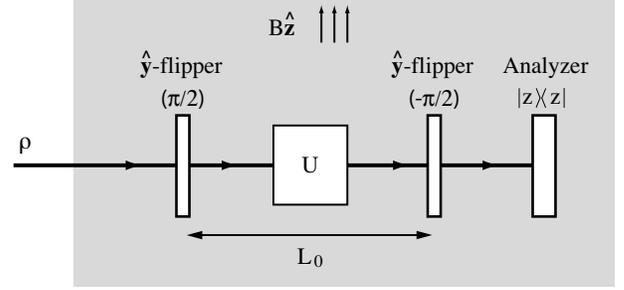}
\caption{Conceptual view of the experimental setup for measuring 
noncyclic phases with polarimetry. Spin$-\frac{1}{2}$ particles 
polarised in the $+z$ direction and carrying a magnetic 
moment $\mu$ are sent through an SU(2) unitarity, surrounded by two 
$\pi /2$ spin flippers. By rigid translation of the spin flippers 
at relative distance $L_0=n\pi v/|\mu B|$, $n$ integer and $v$ 
the particle speed, the noncyclic mixed state phase is extracted 
from the output intensities registered at the analyser.}
\end{figure}

To extract the Pancharatnam relative phase and visibility in a
noncyclic spinor evolution, an extra variable phase shift
$\pm\frac{1}{2}\eta$ must be applied to $\ket{\pm z}$. If the
particles carry a magnetic moment $\mu$, this extra phase shift 
$\eta$ could be implemented by a guiding magnetic field 
$B\hat{{\bf z}}$ put over the entire setup and the variation 
of $\eta$ is achieved by translating the pair of flippers, 
keeping their relative distance $L_0$ constant. By choosing 
$L_0=n\pi \hbar v/|\mu B|$, $n$ integer and $v$ the particle 
speed, one obtains in the pure state case the output intensity 
\cite{wagh95,remark1} 
\begin{equation}
I=\cos^2 \xi \cos^2 \delta + \sin^2 \xi \sin^2(\zeta-\eta) 
\label{eq:pureintensity}
\end{equation} 
with SU(2) parameters $\xi ,\delta, \zeta$ defined by 
\begin{eqnarray}
U & = & e^{i\delta} \cos \xi \ket{+z} \bra{+z} + 
e^{i\zeta} \sin \xi \ket{-z} \bra{+z}
\nonumber \\ 
 & & - e^{-i\zeta} \sin \xi \ket{+z} \bra{-z} 
+ e^{-i\delta} \cos \xi \ket{-z} \bra{-z} . 
\label{eq:su2}
\end{eqnarray}
This yields the extreme values 
\begin{eqnarray}
I_{\textrm{\small{min}}} & = & \cos^2 \xi \cos^2 \delta , 
\nonumber \\ 
I_{\textrm{\small{max}}} & = & \cos^2 \xi \cos^2 \delta + 
\sin^2 \xi 
\end{eqnarray}
upon translation of the flippers. Now, up to a sign, the pure state 
Pancharatnam relative phase $\phi \equiv \arg \bra{+z} U \ket{+z} = 
\delta + \arg \cos \xi$ may be obtained modulo $\pi$ as 
\begin{eqnarray}
\cos^2 \phi & = & \cos^2 \big[ \delta + \arg \cos \xi \big] = 
\cos^2 \delta 
\nonumber \\ 
 & = & \frac{I_{\textrm{\small{min}}}}{1-I_{\textrm{\small{max}}} + 
I_{\textrm{\small{min}}}} , 
\label{eq:wrpanch} 
\end{eqnarray} 
where we have used that $\arg \cos \xi$ is an integer multiple of
$\pi$. Similarly, the pure state visibility $\nu \equiv |\bra{+z} U
\ket{+z} | = |\cos \xi |$ reads
\begin{equation}
\nu = \sqrt{1-I_{\textrm{\small{max}}}+I_{\textrm{\small{min}}}} . 
\label{eq:wrvisibility}
\end{equation}
For a cyclic evolution $(\nu = 1)$ where $\xi = 0$, we have 
$I = \cos^{2} \delta$ and there is no need to translate the 
flippers to obtained the desired phase. When the spinor evolves 
into an orthogonal state corresponding to $\xi = \pi /2$ we have 
$I_{\textrm{\small{min}}} = 0$ and $I_{\textrm{\small{max}}} = 1$ 
so that $\nu = 0$ and $\phi$ is undefined. 

We notice that if $U$ is parallel transporting, the Pancharatnam phase
can be identified with the noncyclic geometric phase. In such a case
$\phi = -\frac{1}{2}\Omega$, $\Omega$ being the solid angle enclosed
by the path and its shortest geodesic closure on the Bloch sphere. For
example, such a parallel transporting unitarity could be realised by a
sequence of SU(2) transformations along great circles on this sphere.

Next, consider a beam of spin$-\frac{1}{2}$ particles with any 
degree of spin polarisation along the $+z$ axis and passing through 
the setup in Fig. 1. Such an input spin state is described by the 
density operator  
\begin{equation}
\rho = \frac{1}{2} \big( 1 + r\sigma_z \big) 
\end{equation}
with $\sigma_z = \ket{+z}\bra{+z} - \ket{-z}\bra{-z}$ and $0\leq r
\leq 1$ being the degree of polarisation (pure states have $r=1$ 
and the maximally mixed state has $r=0$). To compute the output mixed
state intensity $I^{\rho}$ we first notice that the pure state
component $\ket{-z}\bra{-z}$ of $\rho$ gives rise to the 
contribution $1-I$ to the intensity, yielding   
\begin{equation}
I^{\rho} = \frac{1+r}{2} I + \frac{1-r}{2} \big( 1-I \big) = 
\frac{1-r}{2} + rI . 
\label{eq:totalmixedintensity}
\end{equation}
Clearly, $I^{\rho}$ reduces to the pure state intensity 
in Eq. (\ref{eq:pureintensity}) for $r=1$.

The relative phase and visibility for the mixed state $\rho$ 
undergoing the SU(2) evolution described by $U$ in Eq. (\ref{eq:su2}) 
is given by the weighted sum of pure state phase factors according to 
\cite{sjoqvist00}
\begin{equation} 
{\cal V} e^{i\Phi} = \frac{1+r}{2} \cos \xi e^{i\delta} + 
\frac{1-r}{2} \cos \xi e^{-i\delta} .   
\end{equation}
Explicitly, one obtains 
\begin{eqnarray}
\Phi & = & 
\arctan \big[ r\tan \big( \delta + \arg \cos \xi \big) \big] , 
\nonumber \\ 
{\cal V} & = & \nu \sqrt{\cos^2 \delta + r^2 \sin^2 \delta} ,  
\label{eq:mixedppv}
\end{eqnarray}
which reduce to the expected $\delta$ and $\nu$, respectively, 
in the pure state limit $r=1$. Putting the parallel transport 
condition also on $\ket{-z}$ for a geodesically closed path that 
encircles the solid angle $\Omega$, yields the corresponding 
mixed state geometric phase \cite{sjoqvist00} by the 
identification $\delta + \arg \cos \xi = -\frac{1}{2} \Omega$.  

Now, to measure the mixed state phase $\Phi$ and visibility ${\cal V}$
polarimetrically we first notice that 
\begin{equation}
\label{esphase}
\cos^{2} \Phi = \frac{1}{1+r^{2}\tan^{2}\delta} .
\label{eq:mixcos}
\end{equation}
For $r\geq 0$, the extreme values of $I^{\rho}$ in Eq. 
(\ref{eq:totalmixedintensity}) read 
\begin{eqnarray} 
I_{\textrm{\small{min}}}^{\rho} & = & 
\frac{1-r}{2} + r \cos^2 \xi \cos^2 \delta  
\nonumber \\ 
I_{\textrm{\small{max}}}^{\rho} & = & 
\frac{1-r}{2} + r \big[ \cos^2 \xi \cos^2 \delta + 
\sin^2 \xi \big] . 
\end{eqnarray}
Eliminating $\xi$ and inserting into Eq. (\ref{eq:mixcos}), 
we obtain 
\begin{eqnarray}
\cos^{2}\Phi = 
\frac{\big[ I_{\textrm{\small{min}}}^{\rho} - \frac{1}{2}(1-r) \big]/r}
{r \big[ \frac{1}{2}(1+r) - I_{\textrm{\small{max}}}^{\rho} \big] + 
\big[ I_{\textrm{\small{min}}}^{\rho} - \frac{1}{2}(1-r) \big]/r}. 
\nonumber \\ 
\label{eq:cosphase}
\end{eqnarray}
Similarly upon elimination of $\nu = |\cos \xi|$ and $\delta$ in 
the expression for the mixed state visibility ${\cal V}$ in 
Eq. (\ref{eq:mixedppv}) we obtain 
\begin{equation}
{\cal V} = \sqrt{r\left[ \frac{1}{2}(1+r) - 
I_{\textrm{\small{max}}}^{\rho} \right] + 
\left[ I_{\textrm{\small{min}}}^{\rho} - 
\frac{1}{2}(1-r) \right] /r} . 
\label{eq:mixedv}
\end{equation}
Eqs. (\ref{eq:cosphase}) and (\ref{eq:mixedv}) show how to extract 
the mixed state phase and visibility in \cite{sjoqvist00} using 
the polarimetric setup of \cite{wagh95}. They are consistent with 
the pure state case since by putting $r=1$, they reduce to the 
expressions for the Pancharatnam phase and visibility given in 
Eqs. (\ref{eq:wrpanch}) and (\ref{eq:wrvisibility}), respectively. 

\begin{figure}[htb]
\centering
\includegraphics[width=8 cm]{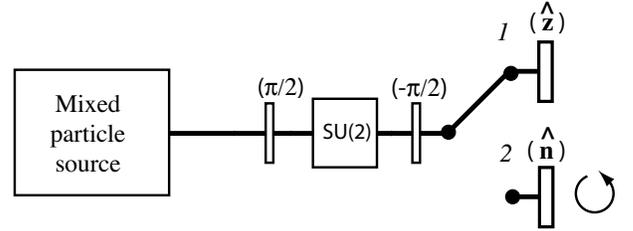}
\caption{Setup to measure the mixed state phase polarimetrically 
without any prior knowledge of the degree of polarisation $r$.  The
SU(2) phase is measured by translation of the spin flippers, while 
$r$ is measured by rotation of the analyser.}
\end{figure}

Let us consider the maximally mixed state case ($r=0$) in some
detail. Here, the density operator is degenerate and the mixed state
geometric phase becomes undefined since no direction is singled out by
the Bloch vector. Yet, the manifestation of the relative phase in
interferometry is a shift of the interference pattern and as such is
well defined also in the $r=0$ case, where it can take the values $0$
or $\pi$. In a polarimetric test, one would obtain
$I_{\textrm{\small{min}}} = I_{\textrm{\small{max}}} = \frac{1}{2}$ in
the maximally mixed state case. This yields $\cos^2 \Phi = 1$
irrespective of the particular form of the SU(2) transformation, which
is consistent with the fact that polarimetry measures phases only
modulo $\pi$. On the other hand, as is apparent from Eq.
(\ref{eq:mixedv}), the experimental extreme intensities
$I_{\textrm{\small{min}}} = I_{\textrm{\small{max}}} = \frac{1}{2}$
for $r=0$ do not determine the visibility ${\cal V}$ uniquely and the
predicted value ${\cal V} = \big| \cos \delta \cos \xi\big|$, obtained
by putting $r=0$ and $\nu = \big| \cos \xi \big|$ in
Eq. (\ref{eq:mixedppv}), cannot be verified. Thus, $r=0$ is a singular
limit in polarimetry.

In experimental situations it seems appropriate to measure $\Phi$
without any prior knowledge of the degree of spin polarisation
$r$. This applies to pure state measurements as well, since in order
to compensate for the reduced degree of polarisation in any real
experimental situation, $r$ must be measured \cite{wagh00}.
Information about $r$ may be obtained by doing a second, distinct
measurement on the system, as sketched in Fig. 2. Suppose we keep 
all previous experimental equipment between the beam source and the
analyser. We then analyse the output intensity
\begin{equation}
\widetilde{I}= \frac{1}{2} + \frac{r}{2} \Big( \big| \bra{{\bf n}} 
\widetilde{U} \ket{+z} \big|^2 - \big| \bra{{\bf n}} \widetilde{U} 
\ket{-z} \big|^2 \Big) 
\end{equation}
in an arbitrary direction $\mathbf{n}$. Here, $\widetilde{U}$ 
denotes the fixed total evolution operator. If the direction 
${\bf n}$ is varied by rotating the spin analyser around some 
suitably chosen axis, the intensity oscillates between the two 
extrema 
\begin{eqnarray}
\widetilde{I}_{\textrm{\small{min}}} & = & \frac{1-r}{2}, 
\nonumber \\
\widetilde{I}_{\textrm{\small{max}}} & = & \frac{1+r}{2} .
\end{eqnarray} 
We may extract $r$ and obtain 
\begin{eqnarray}
\cos^{2} \Phi & = & 
\frac{\Delta I_{\textrm{\small{min}}} /\Delta\widetilde{I}} 
{\Delta\widetilde{I} \Delta I_{\textrm{\small{max}}} + 
\Delta I_{\textrm{\small{min}}}/\Delta\widetilde{I}} , 
\nonumber \\ 
{\cal V} & = & 
\sqrt{\Delta \widetilde{I} \Delta I_{\textrm{\small{max}}} + 
\Delta I_{\textrm{\small{min}}} /\Delta \widetilde{I}} , 
\end{eqnarray}
where we have introduced the semi-positive quantities 
\begin{eqnarray}
\Delta I_{\textrm{\small{min}}} & = & 
I_{\textrm{\small{min}}}^{\rho} - \widetilde{I}_{\textrm{\small{min}}} , 
\nonumber \\ 
\Delta I_{\textrm{\small{max}}} & = & 
\widetilde{I}_{\textrm{\small{max}}} - I_{\textrm{\small{max}}}^{\rho} , 
\nonumber \\ 
\Delta \widetilde{I} & = & 
\widetilde{I}_{\textrm{\small{max}}} - 
\widetilde{I}_{\textrm{\small{min}}} .     
\end{eqnarray} 

To conclude, we have demonstrated that the noncyclic mixed 
state phase discovered in \cite{sjoqvist00} may be tested 
polarimetrically. Such experiments are important as 
high-precision phase tests of unitarily evolving mixed 
states and can for example be realised using partially spin 
polarised neutrons.  
\section*{Acknowledgment}
The work by E.S. was financed by the Swedish Research Council. 


\begin{thebibliography}{99}
\bibitem{remark1} To be more precise, all the intensities in 
this paper are detection probabilities in the output 
channel. The measured intensities are proportional to these 
probabilities with the input intensity as proportionality factor.
\end{thebibliography}
\end{document}